\documentclass[aps,prb,twocolumn,superscriptaddress,longbibliography]{revtex4-1}
\usepackage{hyperref} 
\usepackage{graphicx}
\usepackage{bm}
\usepackage{amsmath}
\usepackage{amssymb}
\begin{document}
\title{Andreev Reflection in Fermi Arc Surface States of Weyl Semimetals}

\author{Yue Zheng}
\affiliation{National Laboratory of Solid State Microstructures and School of Physics, Nanjing University, Nanjing 210093, China}

\author{Wei Chen}
\email{Corresponding author: pchenweis@gmail.com}
\affiliation{National Laboratory of Solid State Microstructures and School of Physics, Nanjing University, Nanjing 210093, China}
\affiliation{Collaborative Innovation Center of Advanced Microstructures, Nanjing University, Nanjing 210093, China}

\author{D. Y. Xing}
\affiliation{National Laboratory of Solid State Microstructures and School of Physics, Nanjing University, Nanjing 210093, China}
\affiliation{Collaborative Innovation Center of Advanced Microstructures, Nanjing University, Nanjing 210093, China}

\begin{abstract}
Fermi arc surface states are the hallmark of
Weyl semimetals, whose identification is
usually challenged by their coexistence with
gapless bulk states. Surface transport measurements
by fabricating setups on the sample boundary
provide a natural solution
to this problem.
Here, we study the Andreev reflection (AR) in a planar
normal metal-superconductor junction on the Weyl semimetal surface
with a pair of Fermi arcs.
For a conserved transverse momentum, the occurrence of normal reflection
depends on the relative orientation between the Fermi arcs and the normal of the
junction, which is a direct result of the disconnected Fermi arcs.
Consequently, a crossover from the suppressed to perfect
AR occurs with varying the orientation of the planar junction,
giving rise to a change from double-peak to plateau structure in conductance spectra.
Moreover, such a crossover can be facilitated by imposing a magnetic field,
making electrons slide along the Fermi arcs so as to switch between
two regimes of the AR. Our results
provide a decisive signature for the detection of Fermi arcs and open the possibilities
of exploring novel phenomenology through their interplay with superconductivity.
\end{abstract}
\maketitle

\section{introduction}

In 1929, Hermann Weyl proposed a new
type of massless fermion with  definite chirality \cite{weyl29pnas}.
After that, great efforts have been made in pursuing such
elemental particles in high-energy particle physics \cite{armitage2018weyl},
yet up until now, no candidate Weyl fermion has been reported \cite{Kajita16rmp,McDonald16rmp}.
In recent years instead, Weyl fermions are surprisingly found in an alternative form
of quasi-particle excitations in a class of solid-state
materials with conic band crossing, called
Weyl semimetals (WSMs) \cite{wan2011topological}.
The discovery of WSM opens up
a new avenue for the study of relativistic
Weyl fermion in condensed matter physics \cite{murakami2007phase,
Burkov11prl,Weng15prx,
huang2015weyl,lv2015experimental,xu2015discovery,
xu2015discovery2,xu2015experimental,xu2016observation,
deng2016experimental,yang2015weyl,huang2016spectroscopic,
tamai2016fermi,jiang2017signature,belopolski2016discovery,
lv2015observation,cpb1}.
It provides an interesting platform for the
experimental testing of predictions made by
quantum field theory
\cite{Adler69pr,bell1969pcac,nielsen1983adler}
in terms of
anomalous transport and optical properties
in the condensed matter context \cite{Zyuzin12prb,Aji12prb,Son13prb,
Chernodub14prb,jian2013topological,burkov2015chiral,
Ma15prb,Zhong16prl,Spivak16prb,hirschberger2016chiral,Huang15prx,shekhar2015extremely,du2016large,Wang16prb,zhang2016signatures,rev1,rev2,rev3,rev4}.

\begin{figure}
\center
\includegraphics[width=0.9\linewidth]{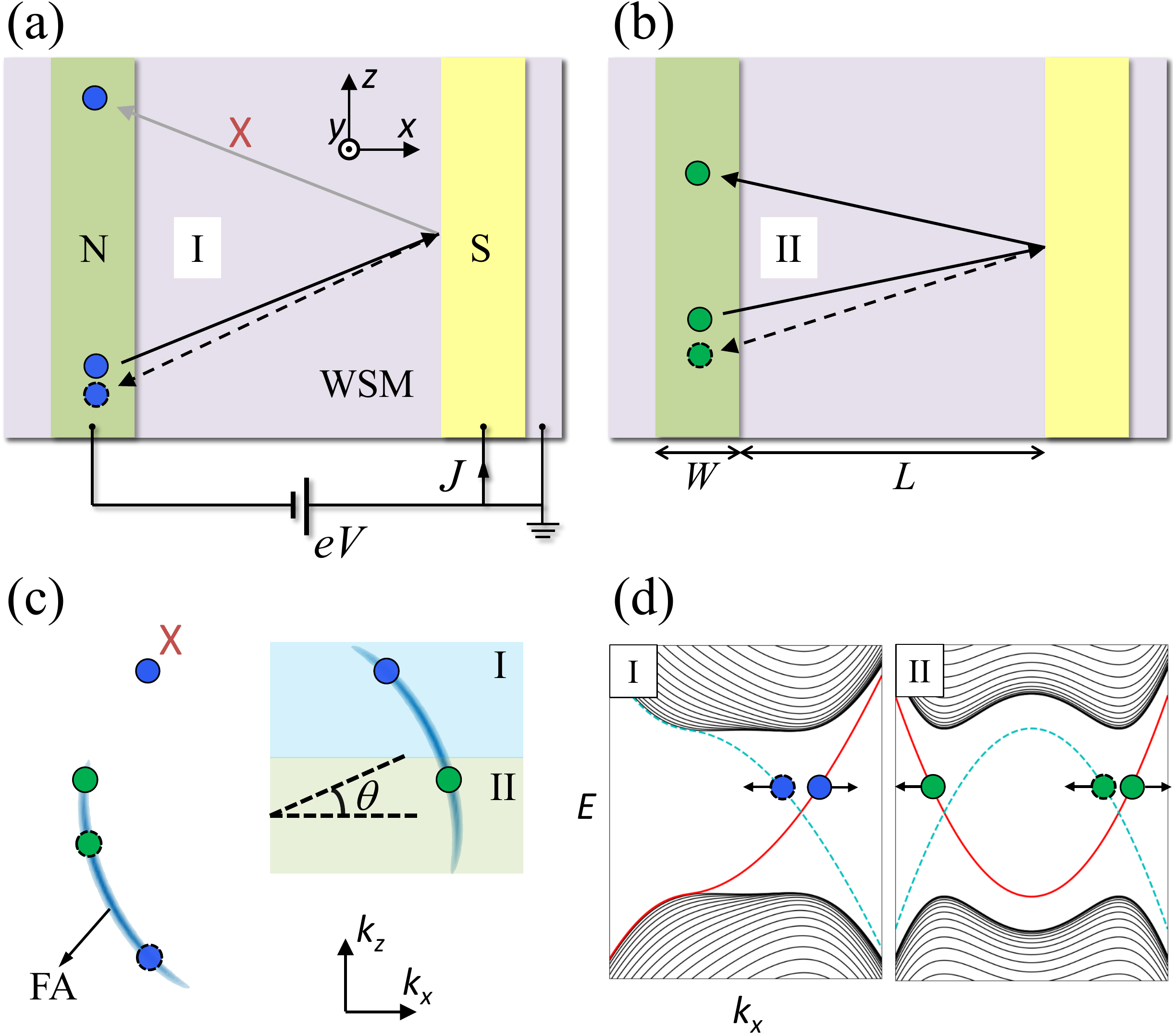}
\caption{(a,b) Schematic of the planar normal metal (N)-superconductor (S)
junction on top of the WSM and the scattering of particles at the interface.
The trajectories of electron (solid circle) and hole (dashed circle) are sketched as
the solid and dashed lines, respectively.
(c) Two regions I and II for AR are
defined by the transverse momentum $k_z$.
(d) Band structures for a fixed $k_z$
in region I and II. The red solid and blue dashed lines
are the electron and hole surface states.}
\label{fig1}
\end{figure}

In addition to its high-energy counterpart,
WSM also exhibits unique properties that exist only
in the condensed matter context, especially the emergence of
Fermi arc (FA) surface states
on the sample boundaries \cite{wan2011topological}.
In WSMs, Weyl points always appear
in pairs with opposite topological charge (chirality) \cite{nielsen1983adler},
the FA spanning between each pair
in the surface Brillouin zone \cite{wan2011topological}.
Such disconnected FAs cannot be
realized in any noninteracting 2D bulk states
so that its emergence can serve as the hallmark of WSMs \cite{huang2015weyl,lv2015experimental,xu2015discovery,
xu2015discovery2,xu2015experimental,xu2016observation,
deng2016experimental,yang2015weyl,huang2016spectroscopic,
tamai2016fermi,jiang2017signature,belopolski2016discovery,
lv2015observation,cpb1}.
It was recently shown that
the configurations of the FA
are sensitive to the details of the sample boundary \cite{morali2019fermi,yang2019topological,Ekahana20prb},
which opens the possibility for engineering FA
and exploring its applications through surface modification.
With this prospect, it is of great importance to extract and analyze
clear signatures of the FA from measurement information in the presence of gapless bulk states.
Surface transport measurements provide a natural solution to this goal
because the setups contact directly to the sample boundaries \cite{chen2018proposal,model}.

In this work, we propose that
Andreev reflection (AR)
in a planar normal metal (N)-superconductor (S) junction
on the WSM surface
can provide a unique signature of the FA.
The junction consists of two parallel
N and S strip electrodes mediated
by the topological surface states in between; see Fig. \ref{fig1}(a).
For the unclosed FA,
there generally exist two regions of different AR scenarios,
referred to as I and II in Fig. \ref{fig1}(c).
In region I, no normal reflection channel is available [Fig. \ref{fig1}(a)]
so that there is a perfect AR
with probability equal to unity within the energy gap [cf. Fig. \ref{fig2}(c)],
in spite of the interface barrier.
On the contrary,
AR is strongly suppressed in region II by
normal reflection at the boundary of the S [Fig. \ref{fig1}(b)]
which results in a pair of resonant peaks
of the AR probability at the gap edges
[cf. Fig. \ref{fig2}(a)].
The proportion of electrons in two regions relies on
the relative orientation between the WSM
and the normal
of the planar junction.
This leads to a crossover between the
double-peak and plateau structures
in the conductance spectra by changing
the orientation of the planar junction
in different samples [Fig. \ref{fig2}(d)].
Two limiting cases occur as the electrons reside entirely in region
I or II [Figs. \ref{fig2}(a) and \ref{fig2}(c)].
Remarkably, such a crossover can be greatly facilitated
by imposing a magnetic field
to a properly orientated planar junction
such that the two AR regions coexist.
The magnetic field drives electrons sliding along
the FA and simultaneously opens up a transport channel
in the bulk, i.e., the chiral Landau band, connected with the surface FA
\cite{potter2014quantum}.
As a result, part of the electrons can
switch between the two AR
regions while the remaining ones
penetrate the bulk with negligible
contribution to the surface transport signature.
In this way, the same crossover
behavior of the conductance
can be achieved
along with a reduction
of its magnitude. The AR spectra have the advantage that two scenarios can be clearly revealed by different shapes of the conductance rather than its magnitude, which provides a distinctive and robust signature of the FAs.

The rest of this paper is organized as follows. In Sec. \ref{model},
we introduce the model and calculation details.
In Sec. \ref{cond}, we show that a crossover of the
conductance from the
suppressed to perfect AR can be achieved by varying the
orientation of the planar junction.
We show that such an effect
can be facilitated by the magnetic field in Sec.
\ref{mag}. Finally, some remarks are given in Sec. \ref{dis}.

\section{model and calculation details}\label{model}

We consider a WSM with four Weyl points,
which can be captured by the effective two-band model as \cite{model}
\begin{equation}\label{H}
\begin{split}
H^0_W(\bm{k})&=M_1(k^{2}_{1}-k^{2}_{x})\sigma _x  + v_yk_{y}\sigma_{y}
+ M_2(k^{2}_{0}-k^{2}_{y}-k^{2}_{z})\sigma_z,\\
\end{split}
\end{equation}
where $v_y$ is the velocity in the $y$ direction, $k_{0,1}$ and $M_{1,2}$ are
parameters, $\sigma_{x,y,z}$ are the Pauli matrices
in the pseudo-spin space. The two bands are degenerate
at four Weyl points $\bm{k}_W=({\pm}k_1, 0 , {\pm}k_0)$.
The main results of this work
are associated with the relative orientation
between the FAs and the planar junction. We define $\theta$ as
the azimuthal angle between the
symmetry axis of the FA and the
normal of the junction (set to the $x$ axis); see Fig. \ref{fig1}(c).
Experimentally, this can be alternatively realized
by fabricating various planar junctions
along different directions.
The FAs with azimuthal angle $\theta$
correspond to the rotated Hamiltonian
$H_W(\bm{k})=H_W^0(U_y^{-1}\bm{k})$
with
$U_y(\theta)=\left(\begin{smallmatrix}\cos \theta & 0 & -\sin \theta \\
            0 & 1 & 0 \\
            { \sin \theta } & 0 & {\cos \theta }\end{smallmatrix}\right)$
as the rotation operator
around the $y$-axis \cite{chen2013specular}.

\begin{figure*}
\center
\includegraphics[width=1.0\linewidth]{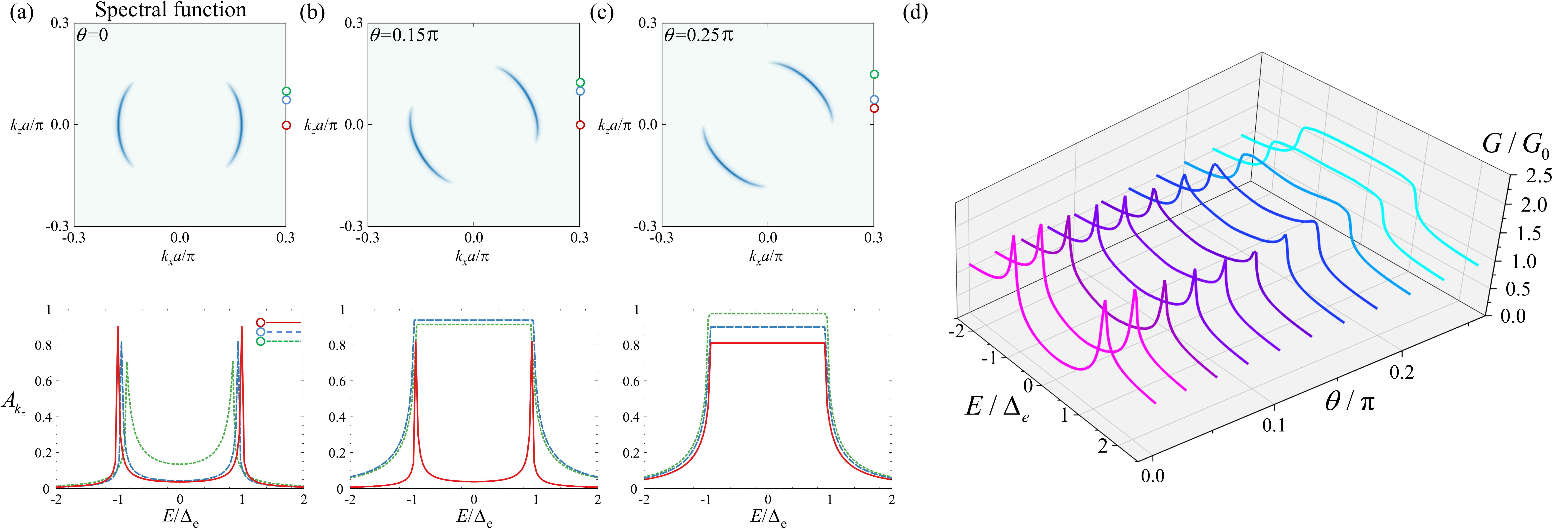}
\caption{(a)-(c) Upper panel: Fermi arc spectra for different azimuthal angle
$\theta$. Lower panel: Corresponding AR probabilities for different $k_z$ channels.
(d) Crossover of the conductance spectra from suppressed to perfect
AR with the variation of $\theta$.
The calculation parameters are $a=1$ nm,
  $M_1=M_2=1.25$ eV nm$^2$, $v_y=-0.66$ eV nm, $k_0=k_1=0.4$ nm$^{-1}$,
$t_N=0.8$ eV, $t_S=0.1$ eV, $C=1$ eV, $\mu_1=2.2$ eV and $\mu_2=2.05$ eV,
$\Delta=5$ meV.
Here, the energy unit is chosen as $\Delta_e=\Delta_e|_{k_z=0}$
for $\theta=0$.
}
\label{fig2}
\end{figure*}

The configurations of the FAs are revealed by
the spectra function $\mathcal{A}(E)=-(1/\pi)\text{Im}G^R(E)$, where $G^R(E)$
is the retarded Green's function,
under the open boundary condition
in the $y$ direction.
To yield curved FAs similar to
those in real materials \cite{Koepernik16prb,belopolski2017signatures,Haubold17prb,Ekahana20prb},
an on-site potential is imposed to the top layer of the WSM lattice, which modifies the dispersion of the surface states\cite{model} through surface band bending effect\cite{curve1,curve2,curve3}
.
The FAs for different azimuthal angle $\theta$
are shown in Figs. \ref{fig2}(a)-\ref{fig2}(c).

To investigate the AR, we write
the Hamiltonian for the whole system in Nambu space.
The electron and hole components are decoupled
in the WSM so that the Bogoliubov-de-Gennes
Hamiltonian takes a diagonal form of
$\mathcal{H}_W(\bm{k})=[H_W^0(\bm{k}), 0; 0, -H_W^{0*}(-\bm{k})]$.
For a given $k_z$, the Hamiltonian $\mathcal{H}_W(\bm{k})$
defines a 2D slice in momentum space.
Edge states induced by the non-trivial band topology of the bulk states emerge as the $k_z$ slice
intersects the FAs
under the open boundary condition \cite{Yang11prb}, which provide scattering channels for electrons (holes).
Throughout the work, all the results are calculated
using a lattice version of this model obtained by substituting $k_{i=x,y,z}\rightarrow a^{-1}\sin k_ia$ and
$k_i^2\rightarrow2a^{-2}(1-\cos k_ia)$, with $a$ the lattice constant of the fictitious cubic lattice (see Appendix \ref{A} for details).

We then discuss the transport process of the 2D slices with different $k_z$. Generally, the FA can be divided into two parts,
according to  whether the normal reflection
channels exist or not; see Fig. \ref{fig1}(c).
In region I, only a single
chiral edge state exists for each $k_z$
so that no normal reflection can occur; see
Fig. \ref{fig1}(d). However, a hole channel
is available for the AR, which corresponds to
the electron-paired state with opposite $k_z$ as sketched by
the blue dashed  circle in Fig. \ref{fig1}(c).
As a result, perfect AR with unity probability
can be realized. Such a band
structure for a given $k_z$  breaks particle-hole symmetry,
which cannot be realized in any 2D system.  It appears
only as a subsystem of the whole 3D WSM, where two
paired electrons are from opposite $k_z$ slices carrying zero
net momentum. As all the $k_z$ channels are taken into account, the particle-hole symmetry retains.
In region II, both normal reflection and AR channels
exist as shown in Figs. \ref{fig1}(c) and \ref{fig1}(d),
similar to a conventional metal.
Given that the normal reflection generally
occurs at the boundary of the S electrode
due to the interface barrier or momentum mismatch,
a suppression of AR is expected in region II, giving rise to
two resonant peaks at the edges of the band gap.

We solve the transport problem through
the surface planar junction as shown in Fig. \ref{fig1}(a).
The N electrode deposited on
top of the WSM is described by the effective Hamiltonian
as $\mathcal{H}_N(\bm{k})=(C\bm{k}^2-\mu_N)\tau_z$
with $C$ determined by the effective mass
of the electron, $\mu_N$ the chemical potential,
and $\tau_{x,y,z}$ the Pauli matrices in Nambu space.
Similarly, the S bar is captured
by $\mathcal{H}_S(\bm{k})=(C\bm{k}^2-\mu_S)\tau_z+\Delta\tau_x$
with a different chemical potential $\mu_S$ and a finite $s$-wave
pair potential $\Delta$.
Due to the proximity effect,
an effective superconducting gap $\Delta_e(k_z)$
can be induced in the FA under the S electrode (see Appendix \ref{B} for details).
We assume that the size of both strip electrodes in the
$z$ direction is much larger than the Fermi wavelength
and their boundaries are smooth enough that
the transverse momentum $k_z$ is approximately conserved
during scattering.
Then by taking $k_z$ as a parameter,
the 3D system can be decomposed into
a set of 2D slices labelled by $k_z$, thus simplifying
the transport calculation.

\section{Conductance}\label{cond}

We first study the transport properties of the 2D slices of the system labelled by $k_z$.
For a hybridized square lattice, the
Hamiltonian at a given $k_z$  is set as $\mathcal{H}_W(\bm{k})$,
$\mathcal{H}_N(\bm{k})$, and $\mathcal{H}_S(\bm{k})$ for different parts.
The coupling between the N (S) and the WSM is described by coupling
strength $t_N$ ($t_S$) between  two outmost lattice layers of contacting areas.
The lattice Hamiltonian for calculation is elucidated in Appendix \ref{A}. The thickness of the WSM and the N (S) electrode in the $y$ direction
is 100 nm and 50 nm, respectively. The width
of the hopping area between the WSM and the N electrode is $W=20$ nm, the
separation between two electrodes is $L=40$ nm [cf. Fig. \ref{fig1}(b)] and
the hopping between the WSM and the S electrode extends to infinity in the
$x$ direction. An on-site potential of $6$ eV is introduced at the boundary line
of the S electrode to simulate the interface barrier or the momentum mismatch
in the planar junction. Both the WSM and N (S) electrodes connect to the leads
extending to infinity in the $\pm x$ directions.
In the energy scale smaller than $\Delta_e(k_z)$,
the current is dominated by the AR. For the two-dimensional lattice with fixed $k_z$, the scattering process can be described by

\begin{equation}\label{rij}
\left[ {\begin{array}{*{20}{c}}
   {\psi _{i,e}^{{\rm{out}}}}  \\
   {\psi _{i,h}^{{\rm{out}}}}  \\
\end{array}} \right] = \sum\limits_j {\left[ {\begin{array}{*{20}{c}}
   {r_{ee}^{ij}} & {r_{eh}^{ij}}  \\
   {r_{he}^{ij}} & {r_{hh}^{ij}}  \\
\end{array}} \right]\left[ {\begin{array}{*{20}{c}}
   {\psi _{j,e}^{{\rm{in}}}}  \\
   {\psi _{j,h}^{{\rm{in}}}}  \\
\end{array}} \right]},\end{equation}
with $\psi^{in(out)}_{i,e(h)}$ represents the income (outgoing) wave amplitudes of electrons (holes) at the N electrode, $r^{ij}_{he}$ describes the scattering amplitude from electrons of channel $j$ to holes of channel $i$ at the same lead.
Then the AR probability can be calculated by taking the trace of the electron-to-hole
reflection matrix, $\text{A}_{k_z}(E)=\text{Tr}[\hat{r}_{he}^{\dagger}(E,k_z)\hat{r}_{he}(E,k_z)]=\sum\limits_{i,j} {|r_{he}^{ij}(E,k_z){|^2}}$,
using KWANT \cite{groth2014kwant},
which describes the AR process that an electron is incident from the
N electrode
and a hole is reflected back.
We plot $\text{A}_{k_z}(E)$ in Figs. \ref{fig2}(a)-\ref{fig2}(c)
for different orientations of the FAs.
In the limiting case of $\theta=0$ in Fig. \ref{fig2}(a),
all $k_z$ is in region II, so that the
AR is strongly suppressed, leaving only two
resonant peaks at $E=\pm\Delta_e$.
Note that $\Delta_e(k_z)$ slightly varies with $k_z$,
so that the positions of resonant
peaks for different $k_z$ do not coincide.
It stems from that the surface states labeled by
$k_z$ possess different spreading in the $y$
direction, which determines the effective
coupling between the surface states
and the superconductor.
Therefore, the proximity effect and the induced gap $\Delta_e(k_z)$
varies with $k_z$; see Appendix \ref{B} for details.
In a general case, e.g., $\theta=0.15\pi$ in Fig. \ref{fig2}(b), region I
and II coexist [cf. Fig. \ref{fig1}(c)].
As $k_z$ lies in region I, perfect AR occurs
with nearly unity probability within $\Delta_e$ ;
In stark contrast, for $k_z$ in region II,
the AR probability exhibits double-spike behavior, which indicates a strong normal reflection.
In the opposite limit, e.g., $\theta=0.25\pi$ in Fig. \ref{fig2}(c),
all the electrons reside in region I, and so the AR probability
 exhibits a perfect AR plateau
within $\Delta_e$. The AR probability is
  slightly smaller than unity which stems from that in our simulation, the
  incident channels of the FA are not fully occupied by the electrons injected
  from the N electrod. Nevertheless, the perfect AR can be manifested in the
  subgap plateau of its probability.

The crossover from suppressed to perfect AR
can be probed by the conductance spectra.
Imposing a bias voltage $V$ between the N
and S electrodes drives a current $J$ in the S; see Fig. \ref{fig1}(a).
The differential conductance
$G=\partial J/\partial V$ at zero temperature can be obtained by
summing the contributions from all the $k_z$ channels as
\begin{equation}\label{G}
G(eV)= \sum_{k_z}g_{k_z};\ \ g_{k_z}(eV) = \frac{{e^2}}{h}(\text{N}_{k_z} + \text{A}_{k_z} - \text{B}_{k_z}),
\end{equation}
where the conductance $g_{k_z}$
for each $k_z$ channel is calculated by the Blonder-Tinkham-Klapwijk
formula \cite{Blonder82prb}. $\text{N}_{k_z}$ is the number of
incident channels below the Fermi energy
in the N electrode and $\text{B}_{k_z}(E)=\text{Tr}[\hat{r}_{ee}^{\dagger}(E,k_z)\hat{r}_{ee}(E,k_z)]=\sum\limits_{i,j} {|r_{ee}^{ij}(E,k_z){|^2}}$
is the normal reflection probability.

The absolute value of conductance $G$ relies on sample
details such as the length of the strip electrodes, which is
not important to our main conclusion. We plot the normalized conductance $G/G_0$
with $G_0=G_{\Delta=0}$ in Fig. \ref{fig2}(d)
for different $\theta$.
The conductance spectra are contributed by all the $k_z$ channels,
thus depending on the weight of two AR regions.
In the limiting case of $\theta=0$,
all the $k_z$ channels are in region II,
giving rise to a double-peak structure in the conductance spectra, the
conductance within the gap being strongly suppressed.
As $\theta$ increases from zero, a portion of $k_z$ channels
transfer from region II to I so that the AR probability with either
double-spike or plateau shape
exists for different $k_z$ channels [Fig. \ref{fig2}(b)].
Consequently, the conductance peaks are lowered
accompanied by a rise of the conductance plateau within the gap.
 As $\theta$ exceeds the threshold $\tan^{-1}(k_0/k_1)$,
all the electrons reside in region I with perfect AR.
Therefore, the conductance exhibits a plateau
within the gap corresponding to perfect AR in all the $k_z$
channels.
The crossover from the double-peak to plateau structure
in the conductance spectra originates from
the high anisotropy of FA configurations
and thus provides a unique signature of the FA.

\begin{figure}
\center
\includegraphics[width=0.9\linewidth]{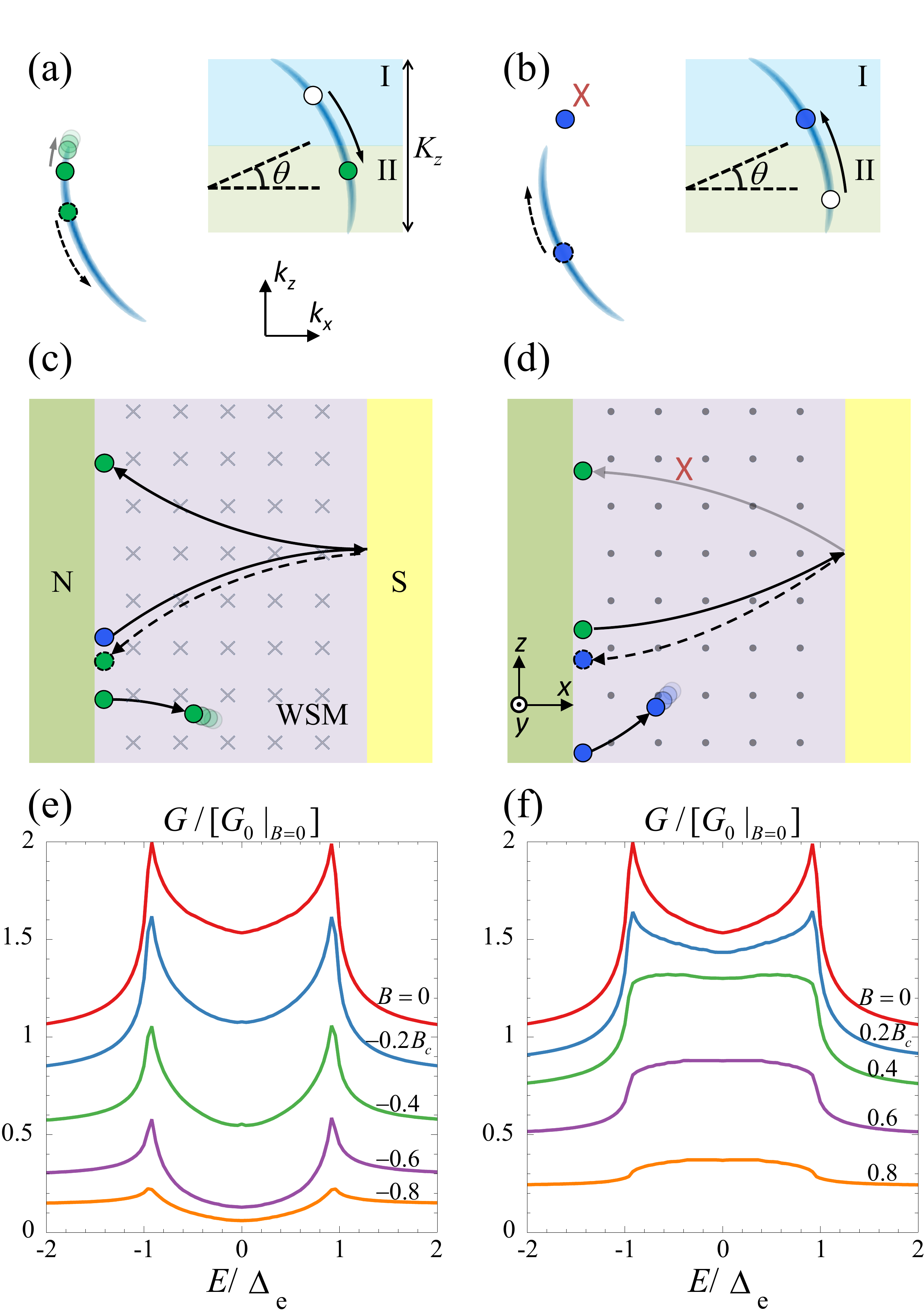}
\caption{(a,b) Particles slide along the FA and switch
between two AR regions driven by
the Lorentz force.
(c,d) Trajectories of particles in real space
corresponding to the upper panels.
Conductances for different magnetic fields in the (e) $-y$
and (f) $y$ directions. All the
parameters are the same as those in
Fig. \ref{fig2}.}
\label{fig3}
\end{figure}

\section{magnetic field effect}\label{mag}

Such a crossover can be more easily observed
by imposing a magnetic field $B$ in the $y$ direction.
In this way, only a single sample
with properly orientated planar junction is required.
Here, we take $\theta=0.15\pi$.
The incident electrons in the right-moving
channels will slide
along the FA by the Lorentz force,
leading to a
switching between two AR regions.
Specifically, for $B<0$, a portion of electrons
originally in region I are driven into region II,
resulting in a transition from perfect
to suppressed AR accordingly; see Fig. \ref{fig3}(a).
Meanwhile, some of the
electrons originally in region II are pushed
into the chiral Landau bands
of the bulk due to the surface-bulk
connection at the Weyl points \cite{potter2014quantum};
see Fig. \ref{fig3}(c).
Those electrons
cannot reach the S electrode so that they do not
contribute to the current $J$ flowing into the S.
On the other hand, some of the reflected electrons
also enter the bulk [cf. Fig. \ref{fig3}(a)]
and they do affect the conductance spectra
via the competition with the AR process.
In short, the magnetic field increases the proportion
of the transport electrons in region II
but reduces the total number of current carriers.
This is well reflected in the conductance spectra of Fig. \ref{fig3}(e),
where with increasing $B$, there are a more visible double-peak structure and
a decrease of the conductance amplitude.
As $B$ exceeds a critical value, all electrons reside in
region II. If $B$ increases further
to the saturation value $B_c=\hbar K_z/(eL)$,
all electrons will transfer into the bulk and there will be no surface electron transport.
Here $K_z$ is the span of the FA in the $k_z$
direction [Fig. \ref{fig3}(a)]
and $L$ is the distance between the N and S electrodes [Fig, \ref{fig1}(b)].
For the parameters $K_z\simeq0.71$ nm$^{-1}$ in Fig. \ref{fig2}(b) and $L=40$ nm,
the saturated magnetic field is evaluated to be $B_c\simeq 11.7$ Tesla (see Appendix \ref{C}
for details).

Similarly, for $B>0$, electrons originally in
region II are driven into region I [Fig. \ref{fig3}(b)], which
induces a transition from the
double-spike to plateau structure of the AR probability.
Consequently, a plateau-like conductance spectrum gradually form
with increasing $B$, accompanied by
a reduction of its overall magnitude due to the transfer
of electrons from region I into
the bulk [Fig. \ref{fig3}(d)]; see Fig. \ref{fig3}(f).
The response of the AR spectra
to the magnetic field stems
from the unique surface-bulk connection
so that it provides another unambiguous evidence
of the FA.

In the calculation, we adopt the Landau
gauge $\bm{A}=({0, 0, B x})$
so that the Peierls substitution
$\bm{k}\rightarrow -i\bm{\nabla}\pm e\bm{A}/\hbar$ (taking $e>0$)
for both the electron and hole parts
retains the $k_z$ conservation.
The number of electrons that
do not reach the S electrode  is subtracted
from $\text{N}_{k_z}$ in Eq. \eqref{G} by
tracking their trajectory in the $-x$ direction
on the bottom surface, and
the number of electrons transferring to the bulk for the
normal reflection is included in $\text{B}_{k_z}$
by tracking their trajectory in the
$x$ direction on the bottom surface as well.
To include the reduction of the
conductance due to the magnetic field,
all results in Figs. \ref{fig3}(e) and \ref{fig3}(f)
are normalized by the same $G_0|_{B=0}$.
For a strong magnetic field $B\simeq B_c$,
all the incident electrons go into the bulk
so that the current $J$
flowing into the S electrode is quenched.

\section{discussion}\label{dis}
Some remarks are made below about the experimental implementation of
our proposal. The surface planar NS junction
can be achieved by state-of-the-art
fabrication techniques \cite{li2020fermi,chen2018finite,ghatak2018anomalous}.
We considered the WSM with
a pair of FAs here, which have been reported in
NbIrTe$_4$ (TaIrTe$_4$) \cite{Koepernik16prb,
belopolski2017signatures,Haubold17prb,Ekahana20prb}, WP$_2$ \cite{Yao19prl},
MoTe$_2$ \cite{Wang16prl},
and YbMnBi$_2$ \cite{borisenko2019time}.
The main conclusion can be generalized to the situation with more Weyl nodes straightforwardly, as the main results stem from the anisotropy of the open FAs and the proximity effect between the superconductor and the FAs, which do not rely on the number of the Weyl nodes.
For the multiple pairs of Weyl nodes, the present results still hold as long as the two regions of the Andreev reflection, without or with backscattering channels [region I and II in Fig. 1(c)], can be well separated in the reciprocal space denoted by $k_z$, then the switching between them can be achieved in the same way by tuning $\theta$ or $B$.
The FA with a regular shape is beneficial
to our proposal, in which the monotonic
change of $k_z$ channels between two AR regions
can be revealed visibly in the conductance spectra.
This requires a big separation
between Weyl points in  momentum space \cite{Sun15prb,Koepernik16prb,
Change16sa,Yao19prl,borisenko2019time,sie2019ultrafast}.
In our calculations, for simplicity, zero chemical potential
was taken in the WSM, where there is a
vanishing density of bulk states.
In real materials
with finite density of bulk states,
our main results remain unchanged
as long as the FAs are well separated from the bulk states
in the surface Brillouin zone.
The presence of bulk states will only cause
certain leakage of surface electrons, but
does not change the current qualitative results.
Finally, we focused on spin-degenerate FA in this work,
and for the FA with fine spin textures \cite{Lv15prl,Xu16prl},
the analysis of AR will be modified by
including the spin degree of freedom.

\begin{acknowledgments}
We thank Xiangang Wan and Feng Tang for helpful discussions.
This work was supported by the National
Natural Science Foundation of
China under Grant No. 12074172 (W.C.), the startup
grant at Nanjing University (W.C.), the State
Key Program for Basic Researches of China
under Grants No. 2017YFA0303203 (D.Y.X.)
and the Excellent Programme at Nanjing University.
\end{acknowledgments}

\begin{appendix}
\section{Lattice model for numerical calculation}\label{A}

We elucidate the model and parameters for
the device sketched in Fig. 1(a).
The numerical calculations are performed on a cubit lattice model of
Eq. \eqref{H} through the mapping $k_{i=x,y,z}\rightarrow a^{-1}\sin k_ia$ and
$k_i^2\rightarrow2a^{-2}(1-\cos k_ia)$, with $a$ the lattice constant of the fictitious cubic lattice.
Performing Fourier transformation in both the $x$ and $y$ directions yields
\begin{equation}\label{Hamiltonian}
\begin{split}
H^{\text{latt}}_{W} = \sum_{i}\psi_i^{\dagger}H_{ii}\psi_i + \sum_{i}\psi_i^{\dagger}H_{i,i+\hat{x}}\psi_{i+\hat{x}} \\
+ \sum_{i}\psi_i^{\dagger}H_{i,i+\hat{y}}\psi_{i+\hat{y}} + H.c.,
\end{split}
\end{equation}
where $\psi_i=(\psi_{1,i}, \psi_{2,i})^{\text{T}}$ are the Fermi operators
with two pseudospin components,
and the on-site and nearest-neighbor hopping matrices are
\begin{equation}
\begin{split}
\begin{aligned}
  &H_{ii}=M_2(k_0^2 - \frac{4}{a^2} + \frac{2}{a^2} \cos k_za)\sigma_z + M_1(k_1^2-\frac{2}{a^2})\sigma_x \\
  &H_{i,i+\hat{x}}=\frac{M_1 \sigma_x}{a^2},\ \ \
  H_{i,i+\hat{y}}=\frac{M_2 \sigma_z}{a^2} + \frac{ v_y\sigma_y}{2ai}.
\end{aligned}
\end{split}
\end{equation}
Note that $k_z$ is conserved during scattering
which is treated as a parameter. Similarly, the
lattice models for the normal metal and superconductor are
\begin{equation}
\begin{split}
\begin{aligned}
  H^{\text{latt}}_{N}&=\sum\limits_i {(\frac{{6C}}{{{a^2}}} - \frac{{2C}}{{{a^2}}}\cos {k_za} - {\mu _N})d_i^\dag {d_i}}\\
                &-\sum\limits_i {\frac{C}{{{a^2}}}} (d_i^\dag {d_{i + \hat{x}}} + d_i^\dag {d_{i + \hat{y}}}) + \text{H.c.} ,\\
  H^{\text{latt}}_{S}&=\sum\limits_i {(\frac{{6C}}{{{a^2}}} - \frac{{2C}}{{{a^2}}}\cos {k_za} - {\mu _S})c_i^\dag {c_i}}\\
                &-\sum\limits_i {\frac{C}{{{a^2}}}} (c_i^\dag {c_{i + \hat{x}}} + c_i^\dag {c_{i + \hat{y}}})
                +\sum\limits_i {\Delta {c^\dag_i}{c^\dag_i}}  + \text{H.c.},
\end{aligned}
\end{split}
\end{equation}
where $d_i$, $c_i$ are electron operators for the normal metal and the superconductor,
respectively. The coupling between the outmost layers of the WSM and the N(S) is described as
\begin{equation}
\begin{split}
\begin{aligned}
 H_T &
  = \sum\limits_i {{t_N}[d_i^\dag {\psi_{1,i + \hat{y}}} + d_i^\dag {\psi_{2,i + \hat{y}}}]} \\
  &+ \sum\limits_i {{t_S}[c_i^\dag {\psi_{1,i + \hat{y}}} + c_i^\dag {\psi_{2,i + \hat{y}}}]}+\text{H.c.}.
\end{aligned}
\end{split}
\end{equation}

\section{Superconducting proximity effect of the surface states}\label{B}

\begin{figure}
\center
\includegraphics[width=0.8\linewidth]{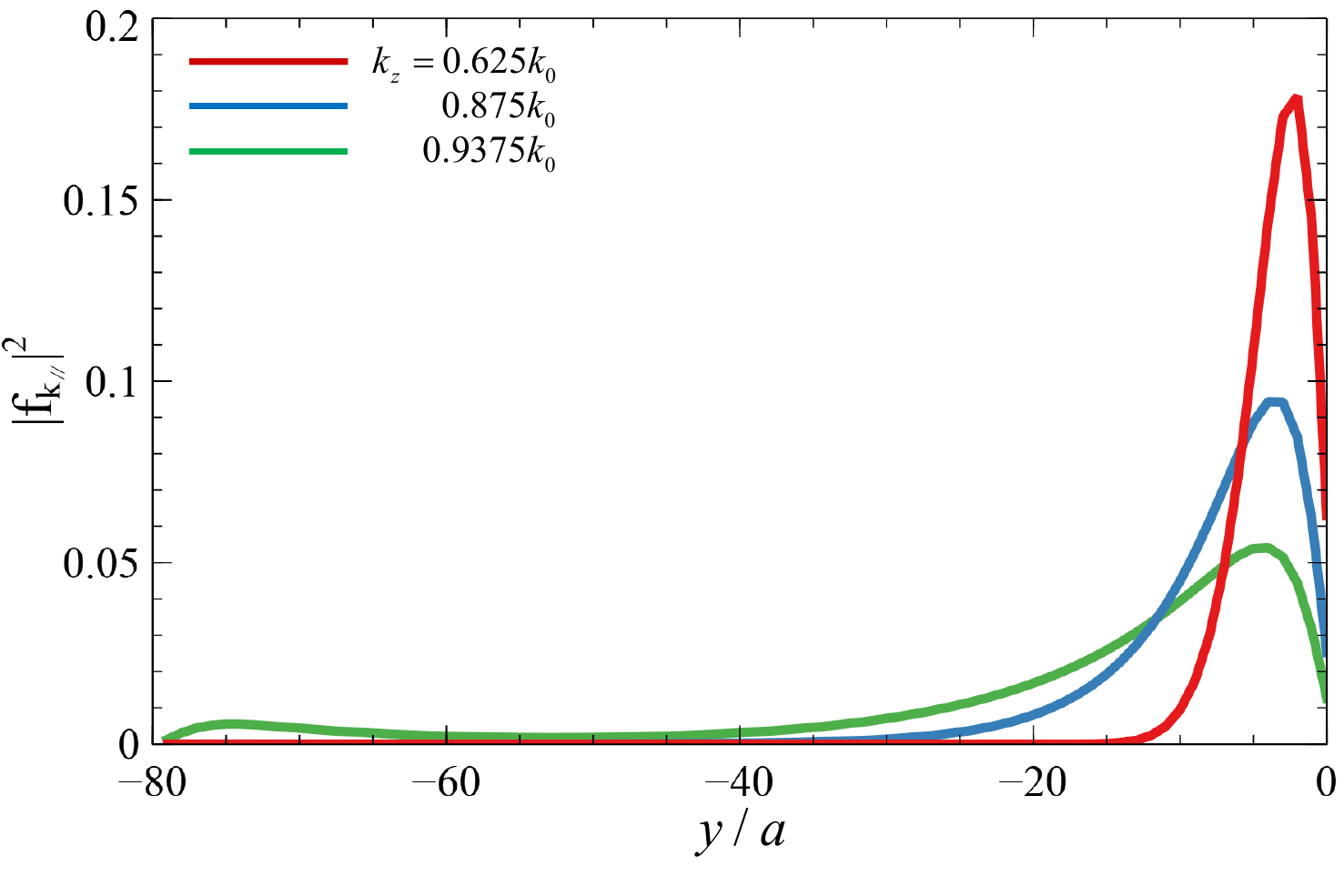}
\caption{The dispersion of $f_{\bm{k}_\parallel}$ along $y$-direction for different $k_z$ channels, here $k_x=k_1$, other parameters are the same as those in
Fig. \ref{fig2}. }
\label{fig4}
\end{figure}

In this section, we employ a tunneling model description
to calculate the effective pair potential $\Delta_e(k_z)$
in the surface states of the WSM induced by
the superconductor deposited above.
We will show that the pair potential $\Delta_e(k_z)$
generally possesses a $k_z$ dependence.
Next we work on the continuous model instead of
the discrete one and the whole Hamiltonian
contains three terms as
\begin{equation}
\begin{split}
H&=H_{S}+H_{W}+H_t,\\
{H_S} &= \sum\limits_{\bm{k}} {{\varepsilon _{\bm{k}}}c_{\bm{k}}^\dag } {c_{\bm{k}}} + (\Delta c_{\bm{k}}^\dag c_{ - {\bm{k}}}^\dag  + H.c.),\\
H_{W}&= \int_{-\infty}^0 dy\sum_{\bm{k}_\|}\psi^\dag_{\bm{k}_\|}(y) H_W^0(\bm{k}_\|,-i\partial_{y})\psi_{\bm{k}_\|}(y),\\
{H_t} &= \sum\limits_{\bm{k}_\|,k_y}\sum_{a=1,2} {\int_{-\infty}^0 {dy\left[ {t(y)c_{\bm{k}}^\dag {\psi _{{{a\bm{k}}_\parallel }}}(y) + \text{H.c.}} \right]} },
\end{split}
\end{equation}
where $H_S$, $H_W^0(\bm{k}_\|,-i\partial_{y})$ [cf. Eq. \eqref{H}] and $H_t$ describe the superconductor,
the WSM and the coupling between them with the strength $t(y)$,
respectively,
and $\psi_{\bm{k}_\|}(y)=[\psi_{1\bm{k}_\|}(y),\psi_{2\bm{k}_\|}(y)]^{\text{T}}$ is the
two-component Fermi operator
in the WSM that is interpreted by the in-plane momentum $\bm{k}_\|=({k_x, k_z})$
and spatial coordinate $y$ in the perpendicular direction.
We assume a good quality of contact between the superconductor and the WSM
such that $\bm{k}_\|=({k_x, k_z})$ is conserved during tunneling.

The coupling between the superconductor and the surface states
strongly relies on the spatial distribution of the latter,
which gives rise to the proximity effect.
Moreover, the pairing occurs mainly around the Fermi level
(here it is the zero energy)
so that it is sufficient to look at the surface states at the Fermi energy.
We solve the surface states for a semi-infinite Weyl semimetal
with the upper surface set to $y=0$. Applying the substitution
$k_y \to -i\partial_y$ to Eq. \eqref{H} and taking $v_y=1$ for simplicity,
the surface states $\phi_{\bm{k}_{\parallel}}(y)$ with zero energy
and $k_z\in(-k_0, k_0)$
is solved by the equation
$H({k_x}, - i{\partial _y},{k_z}) \phi_{\bm{k}_{\parallel}}(y) = 0$,
which gives
\begin{equation}
\begin{split}
\phi_{\bm{k}_{\parallel}}(y)=f_{k_z}(y) \left( {\begin{array}{*{20}{c}}
   \alpha  \\
   \beta  \\
\end{array}} \right),
\end{split}
\end{equation}
with $f_{k_z}(y)={e^{\lambda_1 y}}-{e^{\lambda_2 y}}$
the distribution function in the $y$ direction and ${\lambda _{1,2}}{\rm{ = }}\frac{1}{{2{\rm{|}}{{M}_2}{\rm{|}}}} \pm \sqrt {\frac{1}{{4M_2^2}} + k_z^2 - k_0^2}$.
The spinor $(\alpha, \beta)^{\text{T}}$ is a function of $\bm{k}_{\parallel}$,
with $k_x=\pm k_1$ corresponding to two straight Fermi arcs
solved by the continuous model \eqref{H}.
One can see that $f_{k_z}(y)$ exhibits a $k_z$
dependence, which is also shown in the Fig. \ref{fig4}.
Physically,
the 2D slices labeled by $k_z$ have different mass terms
or gaps in the bulk,
which lead to different spreading of the surface states.
Here only the low energy surface states are of interest
and then $H_W^0$ reduces to
\begin{equation}
\begin{split}
H_{\text{surf}}&=\sum_{\bm{k}_\|}\epsilon_{\bm{k}_\|}\gamma^\dag_{\bm{k}_\|}\gamma_{\bm{k}_\|},\\
\gamma^\dag_{\bm{k}_\|}&=\int dy f_{k_z}(y)[\alpha\psi^\dag_{1\bm{k}_\|}(y)+\beta\psi^\dag_{2\bm{k}_\|}(y)],
\end{split}
\end{equation}
with $\epsilon_{\bm{k}_\|}$ the energy of the surface states for $k_z\in(-k_0, k_0)$,
and $\gamma_{\bm{k}_\|}$ the corresponding Fermi operator.

For the low-energy scale of order $\Delta$, we can interpret the field operator $\psi_{\bm{k}_\parallel}(y)$
by the surface states as
\begin{equation}
\psi_{\bm{k}_\|}(y)\simeq \gamma_{\bm{k}_\|}\phi_{\bm{k}_\|}(y).
\end{equation}
Then the tunneling term reduces to
\begin{equation}
\begin{split}
H_t=\sum_{\bm{k}_\|,k_y} V_{\bm{k}_\|}c^\dag_{\bm{k}}\gamma_{\bm{k}_\|}+\text{H.c.},\\
V_{\bm{k}_\|}=\int dy t(y) f_{k_z}(y)(\alpha+\beta),
\end{split}
\end{equation}
where $V_{\bm{k}_\|}$ is the effective coupling between the surface
states and the superconductor.
It strongly depends on the distribution function $f_{k_z}(y)$
of the surface states and thus on $k_z$.
Starting from the effective tunneling Hamiltonian,
one can solve the self-energy of the surface states
due to its proximity to the superconductor.

The self-energy of the surface states in Nambu space can be
expressed as
\begin{equation}\label{r2}
\begin{split}
{\Sigma_{\rm{surf}}}(\omega ) = {\hat{T}^\dag }{g_S}(\omega )\hat{T},\\
{g_S}({\bm{k}};\omega ) = \frac{{\omega  + {\varepsilon _{\bm{k}}}{\tau _z} + \Delta {\tau _x}}}{{{\omega ^2} - \varepsilon _{\bm{k}}^2 - {\Delta ^2}}},
\end{split}
\end{equation}
where $g_S(\bm{k},\omega)$ is the bare Green's function in the
superconductor and $\hat{T}=V_{\bm{k}_\|}\tau_z$ represents the tunneling terms
from the surface states to the superconductor in $H_t$.
We obtain the self-energy after some algebra as
\begin{equation}\label{r4}
{\Sigma _{\rm{surf}}}({{\bm{k}}_\parallel },\omega ) = |{V_{{{\bm{k}}_\parallel }}}{|^2}
\sum_{k_y} \tau_z g_S(\bm{k},\omega)\tau_z
\end{equation}
Define the 1D density of states as
${N_{{\bm{k}_{\parallel}}}}(\varepsilon ) = {\left[ {\partial \varepsilon /\partial {k_y}} \right]^{ - 1}}$
and take its value approximately to be that at the Fermi energy ${N_{{{\bm{k}}_\parallel }}}(0)$,
which yields
\begin{equation}\label{r5}
\begin{split}
{\Sigma _{\rm{surf}}}({{\bm{k}}_\parallel },\omega ) &=  - \xi (\omega )\left[ {\omega  - \Delta {\tau _x}} \right],\\
\xi (\omega ) &= |{V_{{{\bm{k}}_\parallel }}}{|^2}\pi {N_{{{\bm{k}}_\parallel }}}(0){({\Delta ^2} - {\omega ^2})^{ - \frac{1}{2}}}.
\end{split}
\end{equation}
Then we obtain the full Green's function of the surface states as
\begin{equation}\label{r6}
\begin{split}
{G_{\rm{surf}}}({{\bm{k}}_\parallel },\omega ) &= \frac{\chi(\omega) }{{\omega  - H_{\rm{surf}}^{\rm{eff}}}},\\
H_{\rm{surf}}^{\rm{eff}} &= \epsilon_{\bm{k}_\|}\chi (\omega ){\tau _z} + {\Delta _{e}}\tau_x,\\
\chi (\omega ) &= 1/(1 + \xi)
\end{split}
\end{equation}
where the effective Hamiltonian of the surface states
$H_{\text{surf}}^{\text{eff}}$ involves the proximity effect,
from which we obtains the pairing potential in the surface states as
\begin{equation}
{\Delta _{e}}(k_z,\omega) = \frac{\xi }{{1 + \xi }}\Delta.
\end{equation}
We focus on the weak coupling limit $\xi<<1$
and the effective pairing potential reduces to
\begin{equation}
{\Delta _{e}}(k_z,\omega) \simeq \xi(k_z,\omega)\Delta.
\end{equation}
Note from Eq. \eqref{r5} that $\xi\propto |V_{\bm{k}_\|}|^2$
which increases as $k_z$ deviates from $\pm k_0$,
so that $\Delta_e(k_z)$ also varies with $k_z$,
which explains the slight splitting of the resonant peaks
for different $k_z$ slice in Fig. \ref{fig2}(a).

\begin{figure}
\center
\includegraphics[width=0.8\linewidth]{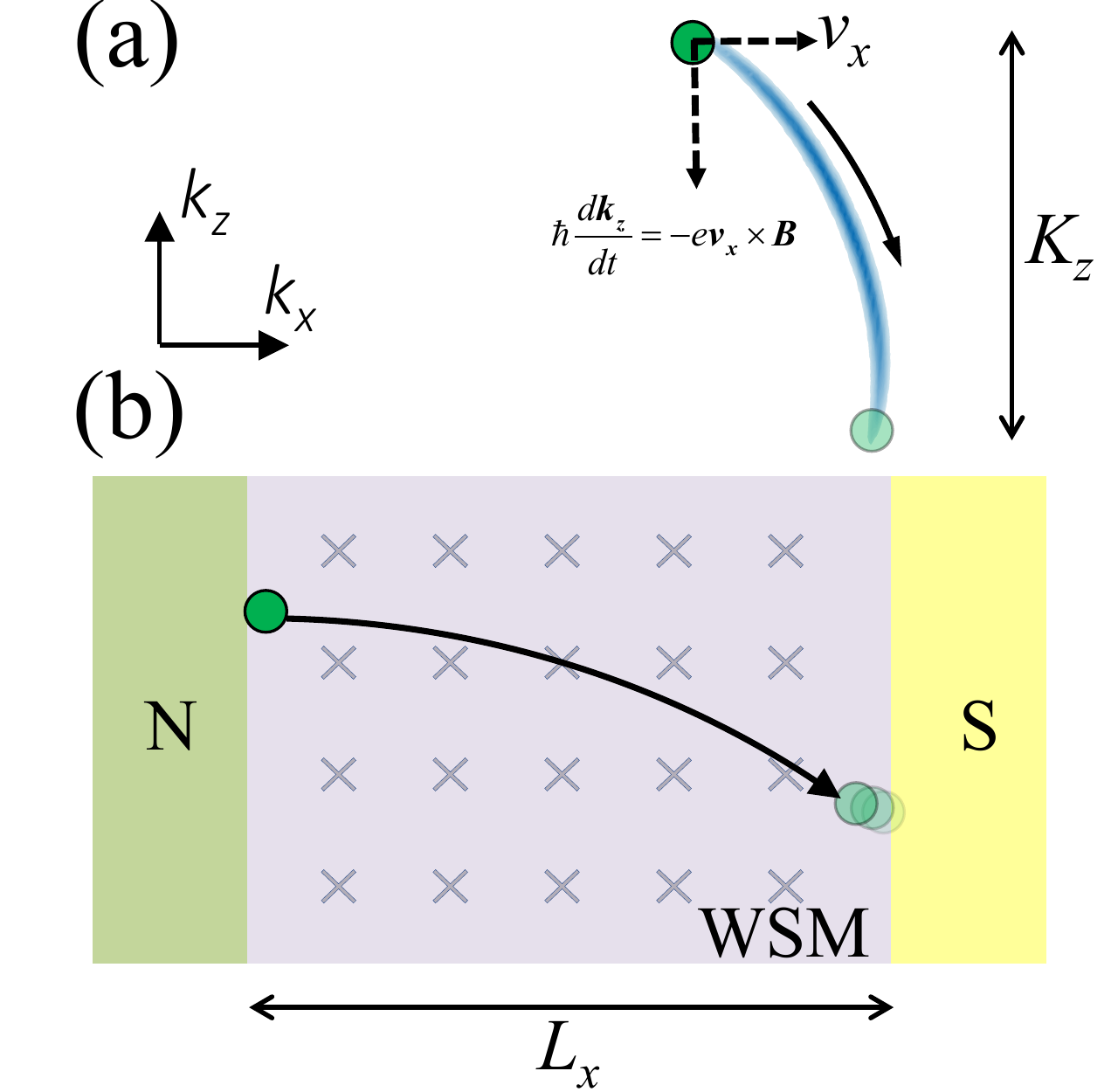}
\caption{(a) Particles slide along the FA and pushed into the Weyl node by the Lorentz force.
(b) The electron trail in real space of the critical case that the incident electrons with $k_z$
at a Weyl node are pushed into the other Weyl node
just before coming out of the magnetic field, corresponding to (a).}
\label{fig5}
\end{figure}

\section{Semiclassical description of the magnetic field effects}\label{C}

In this Section, we evaluate the saturated magnetic field $B_{c}$
based on a semiclassical picture.
As shown in Fig. \ref{fig5}(a), once the magnetic field is employed,
the incident electrons in the right-moving channels
will be driven by the Lorentz force and slide along the
Fermi arc. $B_{c}$ is the critical value such that all
electrons incident from the normal metal arrive the Weyl node and
transfer into the bulk Landau band before
they reach the superconductor. The semiclassical
equation of motion is given by
\begin{equation}
\begin{split}
\hbar \dot{\bm{k}}_z=(-e)\bm{v_{x}}\times \bm{B}
\end{split}
\end{equation}
here $v_x$ is the $x$-direction velocity.
Integrating the equation on both sides yields $\hbar \Delta k_z=-e \Delta x B$,
which relates the change of the momentum $k_z$ in the $z$ direction
and the displacement $\Delta x$ in the $x$ direction.
The saturated magnetic field is thus given by
\begin{equation}
\begin{split}
B_c=\frac{\hbar K_z}{eL_x}.
\end{split}
\end{equation}

\end{appendix}

\end{document}